\documentstyle[aps,prb,epsf]{revtex}
\begin{document}
\twocolumn[\hsize\textwidth\columnwidth\hsize\csname@twocolumnfalse\endcsname
\author{A.E.Koshelev and V.M.Vinokur}
\title{Pinning-induced transition to disordered vortex phase in layered
superconductors}
\address{Materials Science Division, Argonne National Laboratory, Argonne, IL 60439}
\date{\today}
\maketitle

\begin{abstract}
Destruction of the vortex lattice by random point pinning is 
considered as a mechanism of the ``second peak'' transition observed 
experimentally in weakly coupled layered high temperature 
superconductors.  The transition field separating the topologically 
ordered quasilattice from the amorphous vortex configuration is 
strongly influenced by the layered structure and by the nonlocal 
nature of the vortex tilt energy due to the magnetic interlayer coupling.  
We found three different regimes of transition depending on the 
relative strength of the Josephson and magnetic couplings.  The 
regimes can be distinguished by the dependence of the transition field 
on the superconductor parameters and pinning strength.
\end{abstract}
\twocolumn 
\vskip.2pc] 
\narrowtext

\section{Introduction.}

The influence of quenched random pinning on the crystalline order in the
vortex lattice is a subject of longstanding interest. The early pioneering
works by Larkin \cite{Larkin} and Larkin and Ovchinnikov 
\cite{LarkinOvch} introduced the picture of collective pinning in which 
a vortex lattice breaks up into coherently pinned domains, and 
these domains adjust themselves to the pinning field independently of each other.
The relative displacements of vortices grow fast with the distance within
these domains until they reach the characteristic scale of the pinning
potential associated with the vortex core size. The fact that different
Larkin domains are pinned independently suggested that any weak disorder
destroys long range crystalline order. 
The nature of the weakly pinned vortex state was further investigated 
in \cite{mfisher}, where a concept of the low temperature vortex glass 
state with vortices locked near their equilibrium positions by 
infinite pinning barriers had been put forward.  The topological 
nature and the consistency of the elastic description of the vortex 
glass has become a subject of extensive debate.  It was suggested 
\cite{fisher91} that on scales where roughness becomes of the order of 
the lattice constant, topological defects in the form of dislocations 
must appear, converting the elastic vortex glass into a liquid-like 
substance.  This claim was however disputed in \cite{Blatter} where a 
different point of view was advocated, namely, that vortex glass 
induced by weak pinning preserves its elastic integrity on all scales.  
The conclusion of \cite{fisher91} did not account for the fact that on 
scales where vortex lattice roughness becomes of the order of the 
vortex spacing and where according to \cite{fisher91} the generation 
of dislocations was supposed to start, the further growth of the 
roughness slows down and changes from the power-like short-scale 
behavior \cite{Feigelman89} to the logarithmic one \cite{Nattermann}.  
This ultra-slow logarithmic behavior reflects the fact that as the 
roughness exceeds the vortex spacing, the periodic structure of the 
lattice becomes essential and that on the largest spatial scales 
vortex glass maintains quasi-long-range order.  This result of the 
seminal work by Nattermann \cite{Nattermann} was later confirmed and 
developed by Giamarchi and Le Doussal \cite{BraggGlass} and Korshunov 
\cite {korshunov} by using the variational replica method of Bouchaud 
{\it et al}\cite{bouchaud} allowing for derivation of relations
between the different energy and spatial scales within the unique
approach.  Structure factor $S({\bf q})$ of quasilattices with
quasi-long-range order (logarithmic divergence of roughness) shows
power-law Bragg singularities.  Owing to this property Giamarchi and
Le Doussal proposed to call the glassy state resulting from weak
disorder the ``Bragg glass'', to distinguish this state from the
glassy state without any crystalline order.  The possibility of
dislocation-free glassy phase has also been suggested by numerical
simulations of the disordered XY\ model by Gingras and Huse \cite
{GH}.

A vortex lattice close to ideal has indeed been observed in clean 
$Bi_{2}SrCa_{2}Cu_{2}O_{x}$ (BiSCCO) crystals at small fields 
($\lesssim $ 100 Oe) by decorations \cite{lat-dec}, neutron 
diffraction \cite{lat-neut} and Lorentz microscopy \cite{lat-Lor} 
suggesting the existence of the topologically ordered quasilattice at low 
fields.  At higher fields, however, the perfect lattice was never 
observed.  This failure has been taken as the indication that elevated 
magnetic fields promote the action of disorder which destroys the vortex 
lattice.  This view was also supported by numerous transport 
measurements \cite {transport} which show that at high fields the 
resistivity components have long Arrhenius tails indicating 
pinning-dominated behavior.  The conclusion one can draw from the 
above observations is that there should be a disorder-driven phase 
transition between the low-field quasilattice or ``Bragg glass'' phase 
and the high field topologically disordered phase.  Such transition 
indeed has been observed in numerical simulations of Ryu {\it et al.} 
\cite{Ryu} of the model of discrete lines interacting with the 
randomly distributed identical pinning centers.  This understanding 
also correlates well with the recent theoretical findings by Kierfeld 
{\it et al.} \cite{kierf} that the periodic elastic medium should 
remain stable with respect to weak disorder but sufficiently strong 
disorder can trigger the formation of topological defects in the system 
involved.

This putative transition can be identified with the so-called second 
peak field.  A second peak in the magnetic hysteresis loops was 
reported by Kopylov {\it et al.} \cite{Kopylov} in Tl-based HTSC\ 
compounds and by Daeumling {\it et al }\cite{Daeumling} for oxygen 
deficient YBa$_{2}$Cu$_3$O$_{7-x}$ (YBCO).  Further numerous 
investigations 
\cite{Chikumoto91,Chikumoto92,Angadi91,Kadowaki92,Tamegai93,Xu93,Hardy94} 
demonstrated that it is actually a generic feature of anisotropic and 
relatively clean HTSC crystals.  Local magnetization measurements of 
Zeldov {\it et al.} \cite{Zeldov-SecPeak} on high quality BiSCCO 
single crystals reveal that the onset of the peak is very sharp, 
suggesting that the peak indeed may mark some phase transition.  
Smoothness of the peak in earlier global measurement is a consequence 
of spatial average due to inhomogeneous induction inside the samples.  
Field distribution measurements by $\mu ^{+}$SR method \cite{muSR} 
show that vortex line wandering is strongly enhanced above the onset 
of the peak.  Investigations on BiSCCO crystals with different 
oxygen content \cite {Khaykovich,Kishio,Tamegai} and different 
Tl-based compounds \cite {Hardy94} suggest that the peak field scales 
down with the decrease of the interlayer coupling.  However further $\mu 
^{+}$SR studies by Aegerter {\it et al.} \cite{Aegerter} show that in 
BiSCCO compounds with different doping the 
peak field correlates with the London penetration depth $\lambda 
_{ab}$ rather then with the interlayer coupling and scales as $\lambda 
_{ab}^{-2}$.  Controlled irradiation experiments 
\cite{Chikumoto92,Khaykovich1} have shown that the peak field 
decreases with the increase of pinning strength.  These experimental facts 
support the interpretation of the second peak field as a field at which 
the vortex lattice is destroyed by pinning.

A natural question arises: why should the lattice be more ordered at 
smaller fields?  In fact, the collective pinning theory, which for 
many years was the only analytical tool to study pinning phenomena, 
suggests exactly the opposite.  By balancing the elastic and random 
forces, one can immediately see that random forces dominate {\it 
below} some typical field $B_{cp}$ and other vortices cannot prevent a 
given vortex to fall into the local minimum of the random potential at 
$B<B_{cp}$.  Naively, one would expect that the vortex lattice will 
always be disrupted in this individual pinning regime.  This fast 
conclusion occurs not always to be correct.  In the situation when the 
typical distance between local minima is much smaller than the typical 
intervortex spacing, vortices have a wide choice of minima which gives 
them an extra possibility to minimize their interaction energy and 
restore the lattice.  This situation is very typical for HTSC\ due to 
the large difference between the core size of the vortices and the 
intervortex spacing in the experimentally interesting field range.  In 
order to examine vortex lattice stability with respect to weak 
disorder one has to compare elastic and pinning energies stored in an 
elementary ''prism'' of the vortex lattice with the base made by the 
elementary cell and the height equal to the characteristic elastic 
screening length $L_{0}$ along the field direction \cite{LV}, where 
$L_{0}=a_{0}/\gamma $, $a_{0}$ is the lattice spacing, and $\gamma $ 
is the anisotropy parameter.  The analysis of the stability of the 
vortex lattice based on the Lindemann criterion has been done recently 
in \cite{ErtasNelson,Vin,Giamarchi,Kierfeld} for an anisotropic 3D 
superconductor.  This analysis shows that the lattice is expected to 
be stable below some typical field and relates this field with the 
parameters of the superconductor and the pinning strength.  
Unfortunately, most 
of these estimates cannot be applied directly to highly anisotropic 
layered superconductors like BiSCCO, because in 
\cite{ErtasNelson,Giamarchi,Kierfeld} the discrete nature of the vortex 
lines imposed by the layered structure is not taken into account and 
therefore applies only to continuous elastic strings and in \cite 
{Vin} the electromagnetic interaction between vortices was neglected.

In this paper we extend the analysis of 
Refs.~\onlinecite{ErtasNelson,Vin,Giamarchi,Kierfeld} to the case of 
very weakly coupled layered superconductors.  Two important features 
distinguish layered superconductors from three dimensional 
superconductors.  Vortex lines in layered superconductors consist of 
discrete segments (``pancakes'').  All properties of superconductors 
are strongly influenced by this discrete nature of the lines.  For 
continuous elastic strings, which are usually used to model the vortex 
lines, only segments of the order of the Larkin length $l_{c}$ have 
the possibility to explore different minima of the random potential.  In 
layered superconductors the description in terms of elastic strings is 
valid only if $l_{c}$ is much larger than the separation between 
layers $s$.  This condition is violated in the weakly coupled 
layered superconductors like BiSCCO.  When $l_{c}$ drops below $s$, 
pancakes in neighboring layers acquire the possibility to explore 
different minima of the random potential.  Vortex line wandering in 
such situation has been discussed in Ref.\onlinecite{DID} in conjunction with 
the problem of disorder induced decoupling.  
Another important feature of layered compounds is that in the region where 
magnetic interactions between pancake vortices in 
different layers dominate coupling, the tilt energy of the pancake 
stacks becomes strongly non-local, thus modifying significantly the 
overall physical picture of the pinning-induced lattice destruction.

\section{Pinning-induced vortex lines wandering in layered superconductors}

Consider an isolated vortex line in a disordered layered superconductor 
oriented orthogonally to the layers (z-direction).  Adjustment of 
pancake vortices to the pinning potential leads to wandering of the 
vortex line in the z-direction.  The key parameter, which determines 
the amplitude of this wandering, is an elemental wandering distance 
$r_{{\rm w}}$ \cite{DID} or typical distance between two pancakes in 
adjacent layers belonging to the same vortex line.  Two distinct types 
of behavior emerge depending on the relation between the strength of 
the interlayer coupling and the pinning strength.  If the distance 
$r_{{\rm w}}$ is smaller than the core size $\xi $ (weak pinning) the 
discreteness of vortices is not relevant, and their lateral wandering 
is described by the usual elastic string model.  In the opposite 
limit, $r_{{\rm w}}>\xi$, the discreteness plays an essential role.  Now 
pancakes in the neighboring layers have the possibility to stretch their 
relative displacements to distances much larger than the typical 
spacing between the minima of the random potential, to explore a whole 
lot of local minima within the area of $r_{{\rm w}}^{2}$ and thus 
choose the best of all of them.
 
The distance $r_{{\rm w}}$ is determined by the balance between the 
elastic tilt energy and the random energy \cite{DID}.  The tilt energy 
of the vortex line
\begin{equation}
{\cal E}_{{\rm tilt}}[{\bf u}_{n}]=
{\cal E}_{{\rm J}}[{\bf u}_{n}]+{\cal E}_{{\rm M}}[{\bf u}_{n}]
\label{En-tilt}
\end{equation}
consists of the Josephson, 
\begin{equation}
{\cal E}_{{\rm J}}[{\bf u}_{n}]=
\sum_{n}{\frac{\pi E_{{\rm J}}}{{2}}}({\bf u}_{n}-{\bf u}_{n-1})^{2}
\ln {\frac{r_{{\rm J}}}{\left| {\bf u}_{n}-{\bf u}_{n-1}\right| },}  \label{Eel-J}
\end{equation}
and magnetic, 
\begin{equation}
{\cal E}_{{\rm M}}[{\bf u}_{n}]=
\frac{E_{{\rm M}}}{2s}\int \left( dk_{z}\right) \ln
\left( 1+\frac{r_{{\rm cut}}^{2}Q_{z}^{2}}{1+r_{{\rm w}}^{2}Q_{z}^{2}}\right) 
\left| {\bf u}(k_{z})\right| ^{2},  \label{el-M}
\end{equation}
contributions.  Here ${\bf u}_{n}$ is the pancake displacement in the 
$n$-th layer and ${\bf u}(k_{z})$ is the corresponding Fourier 
transform, ${\bf u}(k_{z})=s\sum_{n}\exp (-ik_{z}z_{n}){\bf u}_{n}$.  
$E_{{\rm J}}=\varepsilon _{0}/\left( \pi \gamma ^{2}s\right) $ and 
$E_{{\rm M}}=s\varepsilon _{0}/2\lambda_{ab}^{2}$ are the Josephson 
and magnetic energies per unit area, $\varepsilon _{0}=\Phi 
_{0}^{2}/\left( 4\pi \lambda _{ab}\right) ^{2}$, $r_{{\rm J}}=\gamma 
s$ is the Josephson length, $\gamma =\lambda _{c}/\lambda _{ab}$ is 
the anisotropy ratio of the London penetration depths $\lambda _{ab}$ 
and $\lambda _{c}$, $s$ is the interlayer spacing, and 
$Q_{z}^{2}=\frac{2}{s^{2}}\left( 1-\cos k_{z}s\right) $.  The 
expression under the logarithm in the magnetic coupling tilt energy 
describes the crossovers between different regimes depending on 
the relations between different length scales of the problem 
($\lambda_{ab}$, s, lattice spacing $a$, and $r_{w}$) 
\cite{KoshKes93,Blatter-Magn}. The cut off length $r_{{\rm cut}}$ can 
be estimated from $r_{{\rm cut}}^{-2}\approx \lambda 
_{ab}^{-2}+21.3B/\Phi _{0}$.

Consider a randomly misaligned vortex line with a typical relative 
displacement of adjacent pancakes from one stack $r\sim \left| {\bf 
u}_{n}-{\bf u}_{n-1}\right|$.  The local energy change $\epsilon (r)$ 
per pancake caused by this misalignment includes the loss in the 
coupling energy $\epsilon _{{\rm coup}}(r)$ and the gain in the random 
energy $\epsilon _{{\rm ran}}(r)$, $\epsilon (r)=\epsilon _{{\rm 
coup}}(r)+\epsilon _{{\rm ran}}(r)$.  The coupling term is determined 
by the tilt energy (\ref{En-tilt}) at large wave vectors $k_{z}\sim \pi 
/s$ and, as we mentioned above, can be split into the Josephson and 
magnetic contributions:
\begin{equation}
\epsilon _{{\rm coup}}(r)=
\frac{\pi }{2}E_{{\rm J}}r^{2}\ln \frac{r_{{\rm J}}}{r}
+E_{{\rm M}}r^{2}\ln \frac{r_{{\rm cut}}}{r}.  
\label{coupling}
\end{equation}
Note the different physical origin of these terms: while the loss 
in Josephson energy is related to the interaction of a given pancake 
with pancakes in neighboring layers, the magnetic energy is determined 
by the averaged interaction with pancakes from a large number ($r_{{\rm 
cut}}/s$) of remote layers.  The term $\epsilon _{{\rm ran}}(r)$ 
describes the gain in the pinning energy upon adjusting the position 
of the given pancake to a best minimum of the random potential within 
the area $\pi r^{2}$.  We consider $r\gg \xi $ (weak coupling) so that 
pancakes have the possibility to choose among a large number of minima of 
the random potential.  In such a situation, the function $\epsilon _{{\rm 
ran}}(r)$ is determined by the distribution of the pinning energies, 
which for pinning due to point defects is natural to expect to 
have a Gaussian form.  Accordingly, the concentration of minima with 
energies between $\epsilon $ and $\epsilon +d\epsilon $ is $P(\epsilon 
)d\epsilon $ with
\begin{equation}
P(\epsilon )=\frac{\exp \left( -\frac{\epsilon ^{2}}{U_{p}^{2}}\right)}
{r_{{\rm p}}^{2}U_{{\rm p}}},  
\label{Concentr}
\end{equation}
where $U_{{\rm p}}$ is the pinning energy and $r_{{\rm p}}$ is the typical 
pinning size.  Numerical investigation of random potential for $\delta T_{c}$-pinning give 
$r_{{\rm p}}\approx 6.3\xi $, where $\xi $ is the coherence length.  The 
function $\epsilon _{{\rm ran}}(r)$ is determined by the condition 
that the number of centers with pinning energy less than $\epsilon 
_{{\rm ran}}$ in an area $\sim r^{2}$ is of the order of one, i.e., 
$\pi r^{2}\int^{\epsilon }P(\varepsilon )d\varepsilon \sim 1$.  For 
large displacements, $r\gg r_{{\rm p}}$, this condition gives
\begin{equation}
\epsilon _{{\rm ran}}=-U_{p}\ln ^{1/2}\left( \frac{r^{2}}{2\sqrt{\pi }r_{p}^{2}}\right).  
\label{erandom}
\end{equation}
The optimization of $\epsilon (r)$ with respect to $r$ gives the following
equation for $r_{{\rm w}}$: 
\begin{equation}
r_{{\rm w}}^{2}=\frac{U_{p}}{\left( \pi E_{{\rm J}}\ln \frac{r_{{\rm J}}}{r_{{\rm w}}}
+2E_{{\rm M}}\ln \frac{r_{{\rm cut}}}{r_{{\rm w}}}\right) \ln
^{1/2}\left( \frac{r_{{\rm w}}^{2}}{2\sqrt{\pi }r_{p}^{2}}\right) },
\label{rwander}
\end{equation}
which is valid if $r_{{\rm w}}\gg \xi $.  In the limit where 
the interlayer coupling is not too weak, $\gamma <\lambda _{ab}/s$, 
the distance $r_{{\rm w}}$ is mainly determined by the Josephson 
coupling, and an approximate solution of Eq.~(\ref{rwander}) is given by
\begin{equation}
r_{{\rm wJ}}^{2}\approx \frac{U_{p}}{\pi E_{{\rm J}}\ln \frac{s\varepsilon
_{0}}{U_{p}}\ln ^{1/2}\left( \frac{U_{p}}{E_{{\rm J}}r_{p}^{2}}\right) };
\label{rwandJos}
\end{equation}
In the opposite limit of very weak coupling, $\gamma >\lambda _{ab}/s$, 
the wandering length is determined by the magnetic coupling,
\begin{equation}
r_{{\rm wM}}^{2}\approx \frac{U_{p}}{E_{{\rm M}}
\ln \frac{E_{{\rm M}}r_{{\rm cut}}^{2}}{U_{p}}\ln ^{1/2}
\left( \frac{U_{p}}{E_{{\rm M}}r_{p}^{2}}\right) }.  
\label{rwanMagn}
\end{equation}
In the latter case $r_{{\rm w}}$ is mainly controlled by the London penetration
depth $\lambda _{ab}$ and does not depend upon the anisotropy ratio at all.

Point pinning induces lateral random wandering of the vortex line as 
it traverses a sample in the z-direction so that the line 
displacements grow as ${\bf u}_{n}={\bf u}(z_{n})$, which is commonly 
characterized by the mean squared displacement or {\it roughness}, 
$w(z_{n})$, $w(z_{n})=\sqrt{\left\langle \left( {\bf u}_{n}-{\bf 
u}_{0}\right) ^{2}\right\rangle }$.  The shape of the function 
$w(z_{n})$ is determined by the nature of interlayer coupling.

In the case of dominating magnetic coupling the energy of the mismatched 
pancake stack (\ref{el-M}) is strongly nonlocal, i.e., the energy cost 
of pancake displacement in a given layer is not simply determined by 
the pancake positions in two adjacent layers but it is determined by the 
whole line configurations at distances $r_{\rm cut}=\min(\lambda 
_{ab}, a)$ from the given layer.  A rough estimate for the 
displacement field at small distances $z\ll r_{{\rm cut}}$ can be 
obtained if we neglect weak logarithmic $k_{z}$-dependence of the tilt 
energy (\ref{el-M}) which allows us to describe the collective 
interaction with a large number of pancakes in terms of an effective 
``cage potential''
\begin{equation}
{\cal E}_{{\rm M}}^{{\rm (cage)}}\approx E_{{\rm M}}
\ln \left( \frac{r_{{\rm cut}}}{r_{{\rm wM}}}\right) \sum_{n}{\bf u}_{n}^{2}.  
\label{em-cage}
\end{equation}
In the ``cage'' approximation the correlations between the layers are 
absent and the displacement field $w(z)$ does not increase at all
\begin{equation}
w(z)\approx r_{{\rm wM}},\,\text{at }z< r_{\rm cut}.  \label{m_displ}
\end{equation}
At small distances a weak dispersion of the tilt energy is to be taken into
account perturbatively and gives rise to a weak increase of $w(z)$ 
\begin{equation}
w(z)\approx r_{{\rm wM}}
\left[ 1-\frac{s}{4z\ln \frac{r_{{\rm cut}}}{r_{{\rm wM}}}}\right].  \label{m_displ1} 
\end{equation}
Only at very large distances $z\gg r_{\rm cut}$ the nonlocality becomes 
irrelevant and displacements start to grow again.

The Josephson interlayer coupling provides an additional contribution 
to the tilt energy of a misaligned pancake stack (\ref{Eel-J}) which 
is determined by the interactions between neighboring pancakes in 
adjacent layers.  In the case $\gamma <\lambda _{ab}/s$ the Josephson 
energy dominates at small distances.  Due to the locality of 
interactions the pinning-induced relative displacement $w(z_{n})$ 
grows with $z_{n}$ similar to the elastic string:
\begin{equation}
w(z)=r_{{\rm wJ}}(z/s)^{\zeta }.  \label{displ}
\end{equation}
The rate of growth is determined by the universal wandering exponent 
$\zeta $.  Elaborate numerical analysis \cite{WandExp} gives $\zeta 
\approx 5/8$, which is larger than the naively expected ''random 
walk'' exponent $1/2$.  It is also slightly larger that the value 
$3/5$ suggested by the renormalization group analyses 
\cite{Halpin-Healy} and by a simple scaling reasoning 
\cite{Feigelman89} (see also review \cite{Halpin-Healy95}).  Relative 
displacements for the layered superconductor given by 
Eq.~(\ref{displ}) differ from displacement for weakly pinned vortex 
line in 3D superconductors $w(z)=\xi (z/l_{c})^{\zeta }$ by the 
elemental lengths: $r_{{\rm wM}}$ replaces the core size $\xi$ and the 
interlayer spacing $s$ comes instead of the Larkin length $l_{c}$.  
The 3D expression for $w(z)$, which has been used in Refs.~\onlinecite 
{ErtasNelson,Giamarchi,Kierfeld}, is valid only if $l_{c}>s$.  The last 
condition is strongly violated in highly anisotropic layered 
high-T$_{c}$ materials like BiSSCO and in these materials one has to 
use Eqs.~(\ref{m_displ},\ref{m_displ1}) or Eq.~(\ref{displ}) depending on relative strength 
of the Josephson and magnetic couplings.  

\begin{figure}
\epsfxsize=3.2in 
\epsffile{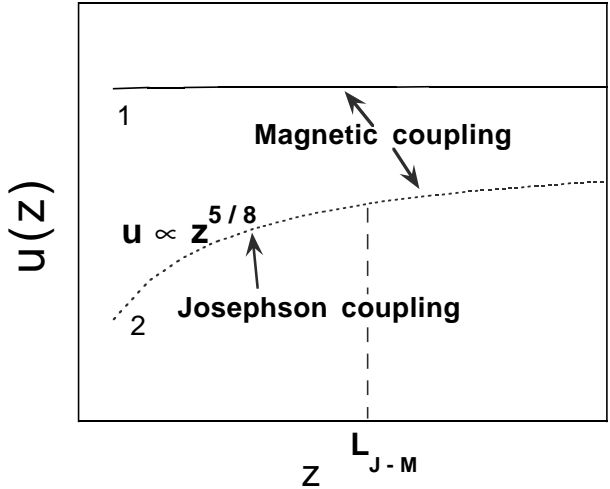}
\caption{Regimes of the vortex line wandering in layered 
superconductors. If only magnetic interlayer coupling is present, 
then displacements don't grow with distance (curve 1). For moderate 
Josephson coupling displacements grow as $z^{5/8}$ at small distances  
and saturate at length $L_{\rm J-M}$ (curve 2)}
\label{Fig-LineWander}
\end{figure}

Note that even in the case of relatively 
strong Josephson coupling ($\gamma <\lambda _{ab}/s$) the growth of 
$w(z)$ at large distances is restricted by magnetic interactions.  
Comparison of the Josephson (\ref{Eel-J}) and magnetic (\ref{el-M}) 
tilt energies shows that above the typical length $L_{\rm J-M}\approx 
\pi \sqrt{\frac{2}{\ln \left( \gamma \right) }}\frac{\lambda 
_{ab}}{\gamma }$ the magnetic coupling takes over again and the 
displacement field crosses over from dependence (\ref{displ}) to 
the almost $z$ independent behavior
\begin{equation}
w(L_{{\rm J-M}})\approx r_{{\rm wJ}}\left( \frac{\lambda _{ab}}{\gamma s}\right) ^{\zeta }.  
\label{wLJ_M}
\end{equation}
Different regimes of line wandering are illustrated in Fig.~\ref{Fig-LineWander}.

\section{Destruction of the vortex crystal by disorder}

\subsection{Interaction length scale $l_{3D}$}

Lateral displacements of an individual vortex line would grow 
infinitely with $z$, however the presence of other vortices restricts 
its wandering.  A given line starts to feel its neighbors at scales 
exceeding some typical length $l_{3D}$, which marks the crossover 
between the single vortex and three dimensional (bundle) regimes.  To 
estimate this length one has to balance the typical tilt energy at the 
wave vector $\pi /l_{3D}$, $C_{44}(\pi /l_{3D})^{2}$ and the typical 
shear energy $4\pi C_{66}B/\Phi _{0}$.  This gives the following 
estimate
\begin{equation}
l_{3D}\approx \sqrt{\frac{C_{44}\Phi _{0}}{C_{66}B}}.  \label{lint}
\end{equation}
The individual line description applies as long as $l_{3D}$ is larger than the
separation between the layers.  The shear and tilt moduli for layered 
superconductor
in the field range $\Phi /\lambda _{ab}^{2}<B\ll H_{c2}$ are given by 
\cite{GK,BrandtSudbo,KoshKes93} 
\begin{eqnarray}
C_{66} &=&\frac{B\Phi _{0}}{(8\pi \lambda _{ab})^{2}};  \label{moduli} \\
C_{44}({\bf k}) &=&\frac{B^{2}/4\pi }{1+\lambda _{c}^{2}k_{\parallel
}^{2}+\lambda _{ab}^{2}Q_{z}^{2}}+\frac{B\Phi _{0}}{2(4\pi \lambda _{c})^{2}}
\ln \frac{k_{\max }^{2}}{K_{0}^{2}+(Q_{z}/\gamma )^{2}} \nonumber\\
&+&\frac{B\Phi _{0}}{2(4\pi )^{2}\lambda _{ab}^{4}Q_{z}{}^{2}}
\ln \left(1+\frac{r_{{\rm cut}}^{2}Q_{z}^{2}}{1+r_{{\rm w}}^{2}Q_{z}^{2}}\right)  \nonumber
\end{eqnarray}
with $K_{0}=\sqrt{4\pi B/\Phi _{0}}$ being the average radius of the 
Brillouin zone.  Three terms in the tilt modulus correspond to the 
nonlocal collective, Josephson single vortex and magnetic 
contributions respectively.  In our case the cutoff wave vector 
$k_{\max }$ in the second term of $C_{44}$ can be estimated as 
$k_{\max }\approx \pi /r_{{\rm w}}$.  The nonlocal tilt modulus in 
Eq.~( \ref{lint}) should be taken at $k_{z}=\pi /l_{3D}$ and 
$k_{\parallel }=K_{0}$.

Two field regimes exist depending upon the strength of the interlayer coupling. For
small Josephson coupling $\gamma >\lambda _{ab}/s$ the length $l_{3D}$
is determined by the magnetic coupling in the whole field range and decays
exponentially at $B>\Phi /\lambda _{ab}^{2}$ 
\begin{equation}
l_{3D}\approx a\exp \left( -\frac{B(\pi \lambda _{ab})^{2}}{4\Phi _{0}}\right) .
\label{l3DMagn}
\end{equation}
At field 
\begin{equation}
B_{{\rm M}cr}\approx \frac{4\Phi _{0}}{(\lambda _{ab}\pi )^{2}}\ln \frac{
\lambda _{ab}}{r_{\rm w}}  
\label{CrossMagn}
\end{equation}
the length $l_{3D}$ matches $r_{\rm w}$, indicating the crossover to 
the quasi-2D regime.

\begin{figure}
\epsfxsize=3.2in 
\epsffile{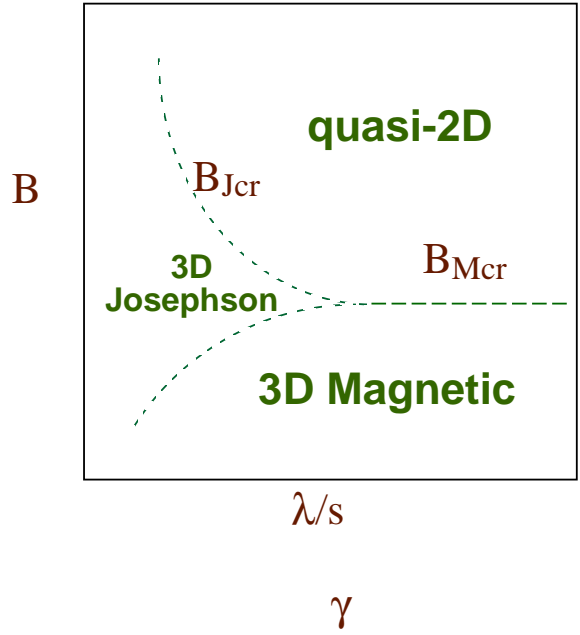}
\caption{Different regimes of lattice behavior in the anisotropy-field plane}
\label{Fig-gam-B}
\end{figure}

In the case of not too weak Josephson coupling $\gamma <\lambda _{ab}/s$,  
two relevant length scales govern the behavior of a single vortex 
line: interlayer spacing $s$ and the crossover length $L_{{\rm 
J-M}}$ introduced in the previous Section.  One can distinguish three 
field regimes corresponding to different relations between $l_{3D}$ 
and these lengths.  At fields smaller than the typical field $B_{{\rm 
J-M}}=\frac{2\Phi _{0}}{(\lambda _{ab}\pi )^{2}}\ln \left( 
\frac{0.1\gamma ^{2}}{\ln (\lambda_{ab}/r_{\rm w})}\right)$ the 
interaction crossover takes place at length scales larger than 
$L_{{\rm J-M}}$ where magnetic coupling still dominates and the length 
$l_{3D}$ is again determined by Eq.~(\ref{l3DMagn}).  At higher fields 
the Josephson coupling dominates and the length $l_{3D}$ is given by
\begin{equation}
l_{3D}\approx 2\frac{a}{\gamma }\sqrt{\ln \frac{a}{r_{{\rm w}}}}.
\label{l3DJos}
\end{equation}
The crossover to the quasi-2D regime takes place at the field $B=B_{{\rm J}cr}$ 
\cite{fisher91,GK}, 
\begin{equation}
B_{{\rm J}cr}<\frac{4\Phi _{0}}{\left( \gamma s\right) ^{2}}\ln \frac{\gamma
s}{r_{{\rm w}}}  \label{CrossJos}
\end{equation}
The schematic field-anisotropy phase diagram shown in 
Fig.~\ref{Fig-gam-B} summarizes the above description of the different 
regimes.

The field at which random pinning destroys the quasilattice or the 
``Bragg glass'' phase can be estimated from the Lindemann-like 
criterion.  \cite {ErtasNelson,Vin,Giamarchi,Kierfeld} The destruction 
of the lattice is expected when the displacement\thinspace $w(z)$ at 
$z\approx l_{3D}$ reaches some fraction of the lattice spacing
  \begin{equation}
      w(l_{3D})=c_{L}a  \label{Criterion}
  \end{equation}
Several regimes exist depending upon the strength of coupling and the strength of
pinning.

\subsection{Dominating magnetic coupling}

In the regime where magnetic coupling dominates ($\gamma >\lambda _{ab}/s$) 
the line roughness $w(z)$ almost does not depend 
on $z$ and the criterion (\ref{Criterion}) can be written simply as 
    \begin{equation}
        r_{{\rm wM}}=c_{L}a,  \label{CriterionM}
    \end{equation}
which leads to the estimate of the field at which the ``Bragg glass'' is
destroyed:
       \begin{equation}
            B_{x}=C_{M}\frac{\Phi _{0}}{\lambda _{ab}^{2}}\frac{T_{m}^{2D}}{U_{p}}
            \label{BxM}
       \end{equation}
Here $T_{m}^{2D}\approx s\varepsilon _{0}/70$ is the melting 
temperature for a single pin-free 2D layer and 
$C_{M}=35c_{L}^{2}\sqrt{\ln \left( \frac{2\lambda 
_{ab}^{2}U_{p}}{s\varepsilon _{0}r_{p}^{2}}\right) }\approx 1\div 2$ 
is a numerical constant weakly depending on the parameters.  The 
melting temperature gives a natural scale for the pinning strength.  
It is important to note that the relative displacement of pancakes in 
neighboring layers becomes comparable with the lattice spacing about 
the same field $B_{x}$.  This means that misalignment transition and 
disorder-induced destruction of the vortex lattice merge and that 
above $B_{x}$ one is going to find the state with both completely 
misaligned and disordered configuration of pancakes.  This is 
very similar to merging of the lattice disordering and the misalignment 
transitions induced by the magnetic coupling first pointed out by 
Blatter {\it et al.} \cite {Blatter-Magn} in the context of the 
melting transition.

A very unusual feature of the transition imposed by dominating 
magnetic coupling is that the transition field (\ref{BxM}) is almost 
insensitive to the shear stiffness of the lattice.  The reason is that 
due to the strong nonlocality of magnetic interactions the wanderings 
of the given line are self-confined, i.e.  the mean squared 
displacements at relevant length scales are mainly determined by the 
interactions of the pancakes belonging to the same string, and 
interaction with other strings very weakly influences these 
displacements.

The disorder-induced destruction of the vortex lattice can occur only 
if $B_{x}$ lies below the dimensional crossover field $B_{{\rm M}cr}$ 
(\ref{CrossMagn}).  This gives the following condition for the pinning 
strength
\begin{equation}
U_{p}\gtrsim 0.5T_{m}^{2D},  \label{CondPin}
\end{equation}
which is simply a condition that pinning is strong enough to destroy 
the 2D lattice in a single layer.  For weaker pinning the ``Bragg 
glass'' remains stable in the quasi-2D regime up to fields close to 
$H_{c2}$.  On the other hand, if $U_{p}$ is substantially larger than 
$T_{m}^{2D}$ then the field $B_{x}$ falls into the region of 
exponentially weak interacting vortices.  In this case the 
intermediate quasilattice state collapses and the vortex lattice is in 
a disordered state throughout the whole field range.  One can conclude 
therefore that the intermediate quasilattice state at fields $\sim 
\Phi _{0}/\lambda _{ab}^{2}$ in magnetically coupled superconductors 
exists only within the limited range of the pinning strength where 
$U_{p}\sim T_{m}^{2D}$.

\subsection{Strong Josephson coupling}

Now we turn to the case $\gamma <\lambda _{ab}/s$.  Due to the 
existence of the two relevant length scales controlling wandering of a 
single line, $s$ and $L_{\rm J-M}$, the behavior is very rich.  Depending 
on the pinning strength the destruction of the ``Bragg glass'' phase 
may take place either in the ``3D Josephson-'' or in the ``3D 
Magnetic'' regimes (see Fig.~\ref{Fig-gam-B}) or it may not happen at 
all.  When the transition falls into the ``3D Josephson'' region we 
obtain from Eqs.(\ref {Criterion},\ref{displ},\ref{l3DJos}) the 
following estimate for the transition field
\begin{equation}
B_{x}=\frac{\Phi _{0}}{\left( \gamma s\right) ^{2}}\left( \frac{c_{w}\gamma
s }{r_{{\rm w}}}\right) ^{2\beta}  \label{Bx3DJos}
\end{equation}
with $\beta={\frac{1}{1-\zeta }}\approx 8/3$ and $c_{w}=c_{L}\left( 4\ln 
\frac{a}{r_{{\rm w}}}\right) ^{-\zeta /2}=0.08 \div 0.1$. Using Eq.(\ref
{rwandJos}) $B_{x}$ can also be connected with the pinning potential $U_{p}$ 
\begin{equation}
B_{x}=\frac{\Phi _{0}}{\left( \gamma s\right) ^{2}}\left( \frac{
C_{J}T_{m}^{2D}}{U_{p}}\right) ^\beta  \label{BxJos}
\end{equation}
with $C_{J}=15\ln \frac{s\varepsilon _{0}}{U_{p}} \left( \ln 
\frac{\Phi_{0}E_{{\rm J}}}{BU_{p}}\right) ^{-\zeta } \sqrt{\ln 
\left(\frac{U_{p}}{E_{{\rm J}}r_{p}^{2}}\right) }c_{L}^{2} \approx 
0.2\div 0.35$.  It is interesting to note that in this regime 
$B_{x}\propto 1/\gamma ^{2}$ in agreement with experimental 
trends \cite{Khaykovich,Kishio,Tamegai}.  Eq.(\ref{BxJos}) is 
valid provided $B_{x}$ falls into the interval 
$B_{\rm J-M}<B_{x}<B_{\rm Jcr}$.  This gives the following conditions for the 
elemental wandering distance
\begin{equation}
0.05\gamma s\lesssim r_{{\rm w}}\lesssim 0.05\gamma s\left( \frac{\pi
\lambda _{ab}}{\gamma s}\right) ^{1-\zeta }.  \label{rw_cond}
\end{equation}
and for the pinning strength 
\begin{equation}
0.5T_{m}^{2D}\lesssim U_{p}\lesssim 0.5T_{m}^{2D}\left( \frac{\pi \lambda
_{ab}}{\gamma s}\right) ^{2(1-\zeta )}.  \label{Up_cond}
\end{equation}
For weaker pinning (smaller $r_{{\rm w}}$) the ``Bragg Glass'' is not
destroyed by pinning. For stronger pinning (larger $r_{{\rm w}}$) the
transition falls into the ``3D Magnetic'' regime. In this regime the
criterion for lattice destruction can be written as. 
\begin{equation}
w(L_{{\rm J-M}})=c_{L}a.
\end{equation}
and leads to the following estimate for the transition field 
\begin{equation}
B_{x}=C_{{\rm JM}}\frac{\Phi _{0}}{\lambda _{ab}^{2}}\left( \frac{\lambda
_{ab}}{\gamma s}\right) ^{2(1-\zeta )}\frac{T_{m}^{2D}}{U_{p}}  \label{BxInt}
\end{equation}
with $C_{{\rm JM}}=70c_{L}^{2}\ln \frac{s\varepsilon _{0}}{U_{p}}\sqrt{\ln
\left( \frac{U_{p}}{E_{{\rm J}}r_{p}^{2}}\right) }\approx 10-15$.

The ``elastic string'' regime of the transition considered in 
Refs.\thinspace \onlinecite{ErtasNelson,Giamarchi,Kierfeld} exists at 
least for some pinning strength only if the crossover from the pancake 
pinning regime to the string pinning regime of an individual vortex line 
occurs at $U_{p}\gtrsim T_{m}^{2D}$.  This gives the relation $E_{{\rm 
J}}\xi ^{2}>T_{m}^{2D}$ which can be rewritten as the condition for 
the anisotropy $\gamma s\lesssim 20\xi $.

\begin{figure}
\epsfxsize=3.2in \epsffile{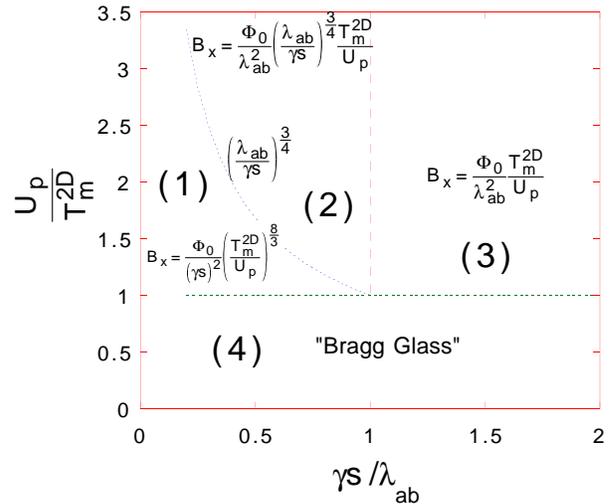} 
\caption{Different regimes of 
disorder driven transition in the pinning strength-anisotropy plane: 
(1) The region of dominating Josephson coupling; (2) The intermediate 
region (Josephson coupling dominates at small scales, magnetic 
coupling dominates at the ``interaction'' length scale); (3) The 
region of dominating magnetic coupling (4) The region of weak pinning 
within which the ordered lattice state remains stable up to the upper 
critical field.  Estimates for the transition fields $B_{x}$ are shown 
for regions (1), (2), and (3)}
\label{Fig-U-gamma}
\end{figure}

Different regimes of quasilattice destruction represented by
Eqs.\thinspace (\ref{BxM},\ref{BxJos},\ref{BxInt}) are summarized in the
anisotropy-pinning phase diagram (Fig.~\ref{Fig-U-gamma}).

\section{Discussion}

Taking typical parameters for optimally doped BiSSCO, 
$\lambda_{ab}\approx 2400\AA$ \cite{Aegerter}, 
$\gamma=200-300$, $s=15\AA$, and $U_{p}=10-15$K we can conclude that 
BiSSCO is located in region (3) of the phase diagram of 
Fig.~\ref{Fig-gam-B} (region of dominating magnetic coupling) but not 
too far from the boundary with region (2) so that the Josephson 
coupling is probably not completely negligible.  Estimate for the 
transition field in region (3) is given by Eq.~(\ref{BxM}).  
Substituting the parameters, we obtain $B_{x}=200-300$ G, in reasonable 
agreement with experiments \cite{Aegerter,Khaykovich}.

The above approach can specify a position for the transition line from the
quasilattice to disordered solid, but can neither prove the existence of
such a transition, nor describe the nature of the resulting high field
entangled solid phase. The first question requiring understanding the
entangled solid state is the very result that enabled us to observe this
transition: what is the mechanism of the sharp increase of the persistent
current measured in the magnetization experiment? Indeed, one can think that
since at the moment of transition pinning in the quasilattice is
dominated by the single vortex pinning regime, the critical current can only
drop with the increase of the magnetic field. However as we have already
mentioned in the Introduction this ``classical'' collective pinning approach
does not account for the possibility that pancakes can explore the deep
bound states. A maximum possible jump of the critical current at the
transition is expected for the case of dominated magnetic coupling. In this
case vortex lines are destroyed at the transition point and it is reasonable
to assume that above the transition each pancake has the possibility to choose
the optimum location within the area $(\pi/4) a_0^2$. If we assume that the
distribution of the energy minima $E_m$ for pancakes obeys Gaussian
statistics $P(E_m)\propto \exp(-E_m^2/U_p^2)$, the typical pinning energy
that a pancake can find on the available area just {\it above} the
transition is estimated as 
\begin{equation}
E_{pin}^{>}\simeq U_p\sqrt{\ln\frac{a_0^2}{\xi^2}}.  \label{crittoka}
\end{equation}
On the other hand, just {\it below} the transition point pancakes has 
the possibility to look for the best pinning center only within the area 
$u^2$, where $u$ is the mean squared deviation of the vortex lines from 
its average position, which at the transition point is given by the 
Lindemann relation $u=c_La$.  Therefore a typical pinning energy {\it 
below} the transition can be estimated as
\begin{equation}
E_{pin}^{<}\simeq U_p\sqrt{\ln\frac{c_L^2a_0^2}{\xi^2}}.  \label{crittokb}
\end{equation}
Since the critical current in a single pancake pinning regime is about 
$J_c\simeq E_{pin}/\xi$, we see that one can expect a substantial 
growth of the critical current at the transition.  For typical 
BiSCCO parameters the critical current can jump by a factor of 2. 
However the experimental jump of the persistent magnetization current is 
only indirectly related to the enlargement of the critical current, because 
in real experiment the persistent current is strongly reduced by 
thermal creep. The jump most probably reflects the enhancement of the 
relaxation barrier above the transition as was indeed observed in 
recent elaborated creep measurements of Konczykowski {\it et al.} 
\cite{koncz}. Quantitative description of this enhancement is a 
challenging problem due to the collective nature of relaxation both 
below and above the transition.

Another important question is whether the entangled solid is a glass or the
glassy state is destroyed by a massive proliferation of dislocations. A very
naive expectation is that dislocations convert the vortex solid into a liquid,
infinite barriers for vortex motion disappear and collective creep over
divergent barriers transforms into a plastic creep governed by the
motion of dislocation over the finite Peierls barriers. More attentive
(although still naive) analysis shows however, that being linear elastic
objects, dislocations may be pinned in turn recovering the glassy
response of the entangled solid. Indeed, recent relaxation measurements on
YBCO crystals \cite{Abulafia95} indicate that creep activation
barriers show the linear current dependence in a wide range of currents
and only at very small currents some upturn in current dependence of energy
is observed, suggesting the appearance of diverging barriers. On the
other hand, it is not quite clear whether this naive dislocation-based
approach applies to highly anisotropic superconductors. One can notice that
neglecting the Josephson coupling one arrives at creep governed by 
two-dimensional pancake diffusion in the entangled phase. Since the
diverging barriers do not appear for 2D particle diffusion, the presence
of electromagnetic coupling only cannot provide glassiness. One can notice
however that switching on even weak Josephson coupling recovers glassy
response, giving rise to a {\it phase glass} (or dislocation glass, or
plastic glass) phase above the entanglement transition. The glassy
properties of the entangled solid will be discussed in detail elsewhere.
Here we point out only that the recent experimental observation on BiSCCO
crystals by Konczykowski {\it et al.} \cite{koncz} reveal a considerable
suppression of the creep rate above the second peak supporting the idea
about the formation of a new dislocation glass in the entangled solid
state.

\section{Acknowledgements}

We would like to thank G.~Blatter, V.~Geshkenbein, M.~Konczykowski,
and E.~Zeldov for numerous fruitful discusions and W.K.~Kwok for
critical reading of the manuscript.  This work was supported from
Argonne National Laboratory through the U.S. Department of Energy,
BES-Material Sciences, under contract No.  W-31-109-ENG-38 and by the
NSF-Office of Science and Technology Centers under contract No.
DMR91-20000 Science and Technology Center for Superconductivity.

\centerline{\bf FIGURES}
\end{document}